\documentclass[prd,preprint,showpacs,showkeys,nofootinbib,floatfix]{revtex4-1}

\usepackage{graphicx}
\usepackage{dcolumn}
\usepackage{bm}
\usepackage{commath}
\usepackage{booktabs}
\usepackage{amsmath,amssymb,amsfonts}
\usepackage{hyperref}

\newcommand{\pom}{{\tt I \! P}}

\begin{document}

\title{Estimations for the Single Diffractive production of the Higgs boson at the Tevatron and the LHC}


\author{M. B. Gay Ducati}

\affiliation{High Energy Physics Phenomenology Group, GFPAE, Instituto de Física, Universidade Federal do Rio Grande do Sul, Caixa Postal 15051, CEP 91501-970, Porto Alegre, RS, Brazil}

\author{M. M. Machado}

\affiliation{High Energy Physics Phenomenology Group, GFPAE, Instituto de Física, Universidade Federal do Rio Grande do Sul, Caixa Postal 15051, CEP 91501-970, Porto Alegre, RS, Brazil}

\author{G. G. Silveira}

\affiliation{High Energy Physics Phenomenology Group, GFPAE, Instituto de Física, Universidade Federal do Rio Grande do Sul, Caixa Postal 15051, CEP 91501-970, Porto Alegre, RS, Brazil}

\affiliation{Center for Particle Physics and Phenomenology (CP3), Universit\'e catholique de Louvain, B-1348 Louvain-la-Neuve, Belgium}

\begin{abstract}
The single diffractive production of the standard model Higgs boson is computed using the diffractive factorization formalism, taking into account a parametrization for the Pomeron structure function provided by the H1 Collaboration. We compute the cross sections at next-to-leading order accuracy for the gluon fusion process, which includes QCD and electroweak corrections. The gap survival probability ($<\!|S|^2\!>$) is also introduced to account for the rescattering corrections due to spectator particles present in the interaction, and to this end we compare two different models for the survival factor. The diffractive ratios are predicted for proton-proton collisions at the Tevatron and the LHC for the Higgs boson mass of $M_H$ = 120 GeV. Therefore, our results provide updated estimations for the diffractive ratios of the single diffractive production of the Higgs boson in the Tevatron and LHC kinematical regimes.
\end{abstract}


\maketitle

\section{Introduction}

Many hard diffractive and high-energy processes are under intense study in the last two decades. One of the main subjects in high-energy physics is the improvement of our knowledge about quantum chromodynamics (QCD). Additionally, the Higgs mechanism is one of most important subjects to be investigated at the LHC, being a cornerstone in the electroweak sector of the standard model (SM). The Higgs boson is expected to be produced by the gluon fusion process ($gg \to H$), making the data analysis of this process an important topic in the project for the LHC experiments, like ATLAS and CMS.

Recent analyses presented an updated estimation of the mass range where it is expected to observe the Higgs boson, which, combining the data coming from CDF and D0 experiments at the Tevatron, have excluded the range $158 < M_{H} < 175$ GeV with $95\%$ of confidence level \cite{:2010ar}. Furthermore, very recent simulations of the ATLAS experimental group have shown that a wider mass range can be excluded with the future LHC data. For instance, with an integrated luminosity of 2 fb$^{-1}$ and a 8 TeV beam energy the range 114 $< M_{H}<$ 500 GeV can be excluded with 95\% confidence level \cite{Collaboration:2010dk}.

The diffractive processes are well described by the Regge theory, where it is considered that a family of resonances is exchanged by the colliding protons \cite{Collins:1977jy}. The leading pole that accounts for this interaction will drive the high-energy behavior of the total cross section, being particularly labeled Pomeron, that has the vacuum quantum numbers \cite{Foldy:1963zz}. However, the nature of the Pomeron is not completely known, as well as its reaction mechanisms, but it is a successful formalism to describe hard diffraction data \cite{Donnachie:1992ny}. Moreover, based on the parton model, it was proposed that the Pomeron could have a partonic content, i.e., quarks and gluons as its constituents, by the Ingelman-Schlein (IS) formalism \cite{Ingelman:1984ns}. Then, systematical observations of diffractive deep inelastic scattering at HERA have increased the knowledge about the Pomeron, providing a diffractive distribution of singlet quarks and gluons into the Pomeron as well as the diffractive structure functions \cite{Aktas:2006hy}.

In this work we are interested in the single diffractive (SD) processes, characterized by the emission of a Pomeron from one of the colliding hadrons that scatters off the other hadron. The cross sections for the SD process are computed at next-to-leading order (NLO) accuracy with QCD and electroweak (EW) corrections, and we use the gap survival probability (GSP) from two different models that accounts for the survival factor for the diffractive Higgs boson production. The cross sections and the diffractive ratios are estimated for the process $p + p(\bar{p}) \to p + H + [LRG] + p(\bar{p})$ for the kinematical regime of the Tevatron ($\sqrt{s}$ = 1.96 TeV) and for those expected to be reached in the LHC ($\sqrt{s}$ = 7, 8 and 14 TeV). In this approach the hard processes will occur by the interaction of the content of one hadron and the content of the Pomeron. In other words, the diffractive cross section is the convolution of the diffractive parton distribution functions (DPDF) and the corresponding partonic cross section, in a similar way as the inclusive case. In addition, diffractive events with a large momentum transfer are also characterized by the absence of hadronic energy in a certain angular regions of the final state, the so-called rapidity gaps. So, the SD processes will present in the final state a large rapidity gap between one proton and the Higgs boson as its main signature.

For the Tevatron kinematical regime, it is known that the data are not correctly predicted with the use of the IS formalism \cite{GayDucati:2007ps,*Kopeliovich:2005ym}, however there are important contributions from unitarity effects to the single-Pomeron exchange cross section that can be considered. These absorptive (unitarity) corrections take into account the fraction of large rapidity gap processes, except elastic scattering, being quite important for the reliability of the predictions for hard diffractive processes. The multi-Pomeron contributions depend on the particular hard process, and one is able to compute the GSP \cite{Chehime:1992bp,*Bjorken:1991xr,*Bjorken:1992er} for a specific production process, which accounts for the fraction of events where the rapidity gaps will be present in the final state after the rescattering events. In this way, the application of a survival factor in the diffractive cross section can correctly describe the high-energy data. For instance, some predictions for $W^{\pm}$, $Z^{0}$, heavy quarks, $\Upsilon$ and $J/\psi$ were presented in Refs.\cite{GayDucati:2007ps,GayDucati:2010vu,*GayDucati:2009rr} for the LHC energies, and it was possible to see that this approach describes very well the Tevatron data.

This paper is organized as follows: in Sec. \ref{inc}, we present the main equations for the inclusive production of the Higgs boson at NLO accuracy. Next, in Sec. \ref{sd-dpe}, we rewrite the parton luminosity in order to introduce the Pomeron exchange from the colliding proton, taking into account the $gg \to H$ production. Further, in Sec. \ref{gsp}, we present the models for the GSP applied in this work, showing the probabilities for each energy regime. Then, in Sec. \ref{results}, we present the estimations for the inclusive and diffractive cross sections as a function of the Higgs boson mass for different collider energies, and also the rapidity distributions of the Higgs boson. Finally, in Sec. \ref{concl}, we summarize our conclusions.

\section{Inclusive production}
\label{inc}

Let us present the main formulas for the inclusive cross sections for the production of Higgs boson in proton-proton collisions. The production process considered in this work is the gluon fusion $pp \to gg \to H$, since it is the leading production mechanism of the Higgs bosons in the high-energy regime \cite{Carena:2002es,*Hahn:2006my,*Duperrin:2008in}. The gluon coupling to the SM Higgs boson is mediated by a triangular loop of quarks, with the leading contribution of the quark top. The production cross section at lowest order is given by \cite{Spira:1995rr}
\begin{eqnarray}
\sigma_{LO}(pp\to H + X)=\sigma_{0}\tau_{H}\frac{\dif{\cal{L}}^{gg}}{\dif\tau_{H}},
\label{equacao34}
\end{eqnarray}
with the Drell-Yan variable defined as $\tau_{H} = M^{2}_{H}/s$, where $s$ is the invariant $pp$ collider energy squared. The gluon-gluon luminosity has the form
\begin{eqnarray}
\frac{\dif{\cal{L}}^{gg}}{\dif\tau}=\int^{1}_{\tau}\frac{\dif x}{x}g(x,M^{2})g(\tau/x,M^{2}),
\label{luminosity}
\end{eqnarray}
with $g(x,M^{2})$ being the gluon distribution function into the proton, where we apply the MSTW2008 parametrization at NLO accuracy for such distribution \cite{Martin:2009bu,*Martin:2009iq}, with $M$ as the factorization scale. In Eq.(\ref{equacao34}), the function $\sigma_{0}$ reads
\begin{eqnarray}
\sigma_{0} = \frac{G_{F} \alpha^{2}_{s}(\mu^{2})}{288\sqrt{2}\pi}\left |\frac{3}{4}\sum_{q}A_{Q}(\tau_{Q}) \right |^{2},
\label{SIGMAzero}
\end{eqnarray}
where $A_{Q}(\tau_{Q}) = 2[\tau_{Q} + (\tau_{Q} - 1) f(\tau_{Q})]/\tau_{Q}^{2}$, and $\tau_{Q} = M^{2}_{H}/4m^{2}_{q}$. In this work it is considered only the leading contribution of the top quark ($m_{q}$ $\equiv$ $m_{t}$ = 172.5 GeV), called heavy-quark limit in Ref.\cite{Spira:1995rr}, and then we are taking the approximation $\tau_{Q} \leq 1$, which means the use of $f(\tau_{Q}) = \arcsin^{2}{\sqrt{\tau_{Q}}}$.

The NLO QCD corrections for the fusion process $gg \to H$ correspond to the processes $gg \to H(g)$, $gq \to Hq$ and $q\bar{q} \to Hg$ \cite{Dawson:1990zj,Spira:1995rr}, introducing virtual and real corrections to the scattering amplitude. The production cross section for the Higgs boson at NLO accuracy in $pp$ collisions is written as \cite{Spira:1995rr}
\begin{eqnarray}
\sigma_{NLO}(pp \to H + X) = \sigma_{0} \left[ 1 + {\cal{C}}\frac{\alpha_s(\mu^{2})}{\pi} \right ]\tau_{H}\frac{d{\cal{L}}^{gg}}{d\tau_{H}} + \Delta\sigma_{gg} + \Delta\sigma_{gq} + \Delta\sigma_{q\bar{q}},
\label{equation38}
\end{eqnarray}
with the renormalization scale in the strong coupling constant $\alpha_{s}$ and the factorization scale in the parton densities to be fixed properly. Particularly, in Eq.(\ref{SIGMAzero}) the strong coupling constant is applied at lowest order accuracy; however, for the NLO contributions the $\alpha_{s}$ is applied at NLO accuracy through the exact numerical solution \cite{Gluck:1998xa}
\begin{eqnarray}
\frac{\dif \alpha_{s}(\mu^{2})}{\dif \ln \mu^{2}} = - \frac{\beta_{0}}{4\pi} \alpha_{s}^{2}(\mu^{2}) - \frac{\beta_{1}}{16\pi^{2}}\alpha_{s}^{3}(\mu^{2}),
\label{nlo_alphas}
\end{eqnarray}
where $\beta_{0} = (11N_{c} - 2N_{F})/3$ and $\beta_{1} = (102N_{c} - 38N_{F})/3$, with $N_{c} = 3$. The $\Lambda$ scale is fixed by the threshold of the quark masses during the $\mu^{2}$ evolution, and fixing the value of $N_{F}$ properly.

The coefficient ${\cal{C}}(\tau_{Q})$ denotes the contributions from two-loop virtual corrections, regularized by the infrared singular part of the cross section for real gluon emission, and is expressed by \cite{Spira:1995rr}
\begin{eqnarray}
{\cal{C}}(\tau_{Q}) = \pi^{2} + c(\tau_{Q}) + \left( \frac{11 N_{c} - 2N_{F}}{6} \right) \log\frac{\mu^{2}}{M^{2}_{H}} ,
\label{equacao39}
\end{eqnarray}
where $\pi^{2}$ refers to the infrared part, and $c(\tau_{Q})$ is a finite function, which, solved analytically, results in $c(\tau_{Q}) = 11/2$ for $\tau_{Q} = M^{2}_{H}/4m^{2}_{q} \ll 1$ \cite{Graudenz:1992pv}.

The $\Delta\sigma_{ij}$ are the hard contributions from gluon radiation in the $gg$ scattering and the $q\bar{q}$ annihilation, and they depend on the renormalization scale $\mu$ and the factorization scale $M$ in the parton densities. These contributions can be expressed by \cite{Spira:1995rr}
\begin{subequations}
\begin{eqnarray}\nonumber
\Delta\sigma_{gg} &=& \int^{1}_{\tau_{H}}\dif\tau\frac{\dif{\cal{L}}^{gg}}{\dif\tau}\frac{\alpha_{s}}{\pi}\sigma_{0} \left \{ -\hat{\tau}P_{gg}(\hat{\tau}){\text{log}}\frac{M^{2}}{s} + d_{gg}(\hat{\tau},\tau_{Q}) \right. \\ &+& \left. 12\left [ \left(\frac{{\text{log}}(1-\hat{\tau})}{1-\hat{\tau}}\right )_{+} - \hat{\tau}[2-\hat{\tau}(1-\hat{\tau})]{\text{log}}(1-\hat{\tau}) \right ] \right\}, \\
\Delta\sigma_{gq} &=& \int^{1}_{\tau_H}\dif\tau\sum_{q,\bar{q}}\frac{\dif{\cal{L}}^{gq}}{\dif\tau}\frac{\alpha_{s}}{\pi}\sigma_{0}\left \{ d_{gq}(\hat{\tau},\tau_{Q}) + \hat{\tau}P_{gq}(\hat{\tau})\left [ -\frac{1}{2}{\text{log}}\frac{M^{2}}{\hat{s}}+{\text{log}}(1-\hat{\tau})\right]\right\}, \\ 
\Delta\sigma_{q\bar{q}} &=& \int^{1}_{\tau_{H}}\dif\tau \sum_{q}\frac{\dif{\cal{L}}^{q\bar{q}}}{\dif\tau}\frac{\alpha_{s}}{\pi}\sigma_{0}d_{q\bar{q}}(\hat{\tau},\tau_{Q}),
\label{equacao40}
\end{eqnarray}
\end{subequations}
where $\hat{\tau} = \tau_{H}/\tau$, and $P_{gg}(\hat{\tau})$ and $P_{gq}(\hat{\tau})$ are the standard Altarelli-Parisi functions \cite{Altarelli:1977zs}
\begin{subequations}
\begin{eqnarray}
P_{gg}(\hat{\tau}) &=& 6 \left\{ \left( \frac{1}{1-\hat{\tau}} \right)_{+} + \frac{1}{\hat{\tau}} - 2 + \hat{\tau}(1 - \hat{\tau}) \right\} + \frac{11N_{c}-2N_{F}}{6} \delta(1-\hat{\tau}), \\
P_{qg}(\hat{\tau}) &=& \frac{4}{3} \frac{1 + (1 - \hat{\tau})^{2}}{\hat{\tau}}.
\label{ap-eq}
\end{eqnarray}
\end{subequations}
The $F_{+}$ denotes the usual $+$ distribution, such that $F(\hat{\tau})_{+} = F(\hat{\tau}) - \delta(1 - \hat{\tau}) \int^{1}_{0}d\hat{\tau}^\prime F(\hat{\tau}^\prime)$. As we are considering the heavy-quark limit, the $d_{ij}$ functions can be solved analytically, resulting in a simpler set of expressions \cite{Spira:1995rr}
\begin{subequations}
\begin{eqnarray}
d_{gg}(\hat{\tau},\tau_{Q}) & = & -\frac{11}{2}(1-\hat{\tau})^{3}, \\
d_{gq}(\hat{\tau},\tau_{Q}) & = & -1+2\hat{\tau}-\frac{\hat{\tau}^{2}}{3}, \\ 
d_{q\bar{q}}(\hat{\tau},\tau_{Q}) & = & \frac{32}{27}(1-\hat{\tau})^{3}.
\label{dfunctions}
\end{eqnarray}
\end{subequations}

\begin{figure}
\includegraphics[width=\textwidth]{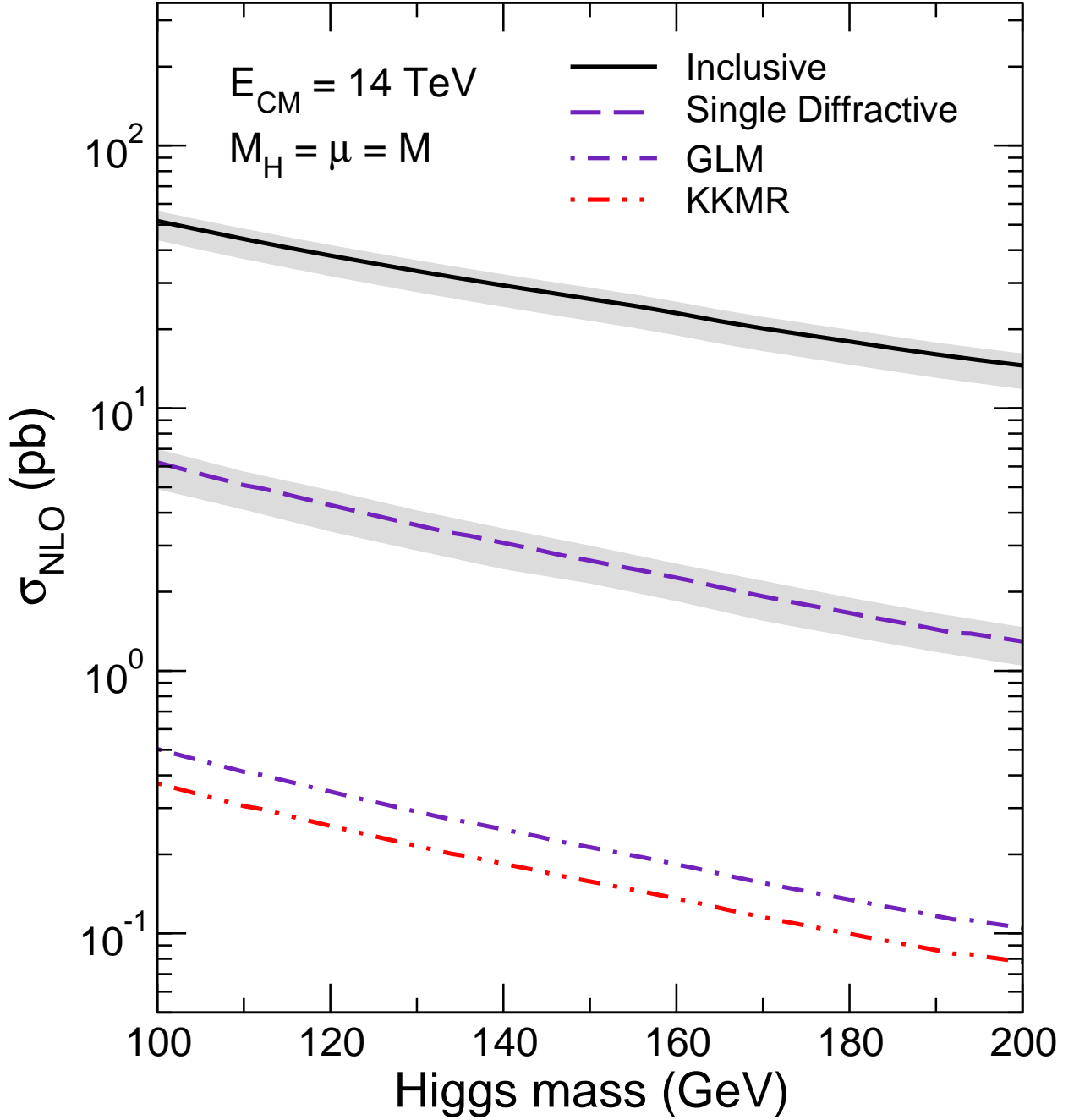}
\caption{\label{fig_lhc} Production cross sections for inclusive and single diffractive processes for the Higgs boson in the LHC. The gray bands show the variation of each cross section for the energy scales $0.5M_{H}<(\mu=M)<4.0M_{H}$. The lower curves show the predictions for the single diffractive production using two different models for the GSP.}
\end{figure}

\begin{figure}
\includegraphics[width=\textwidth]{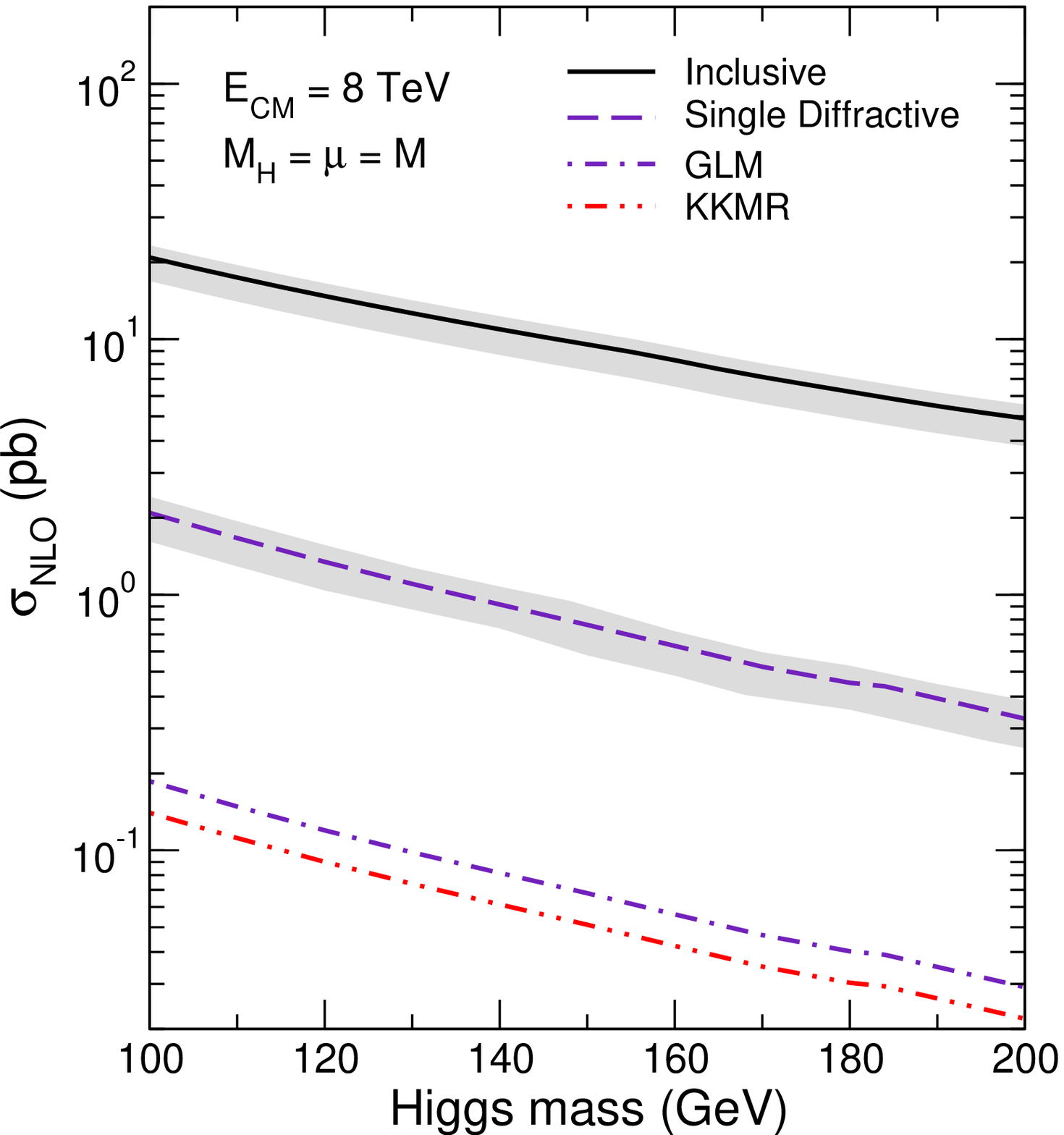}
\caption{\label{fig_8} The same as Fig.\ref{fig_lhc} for $\sqrt{s}$ = 8 TeV.}
\end{figure}

\begin{figure}
\includegraphics[width=\textwidth]{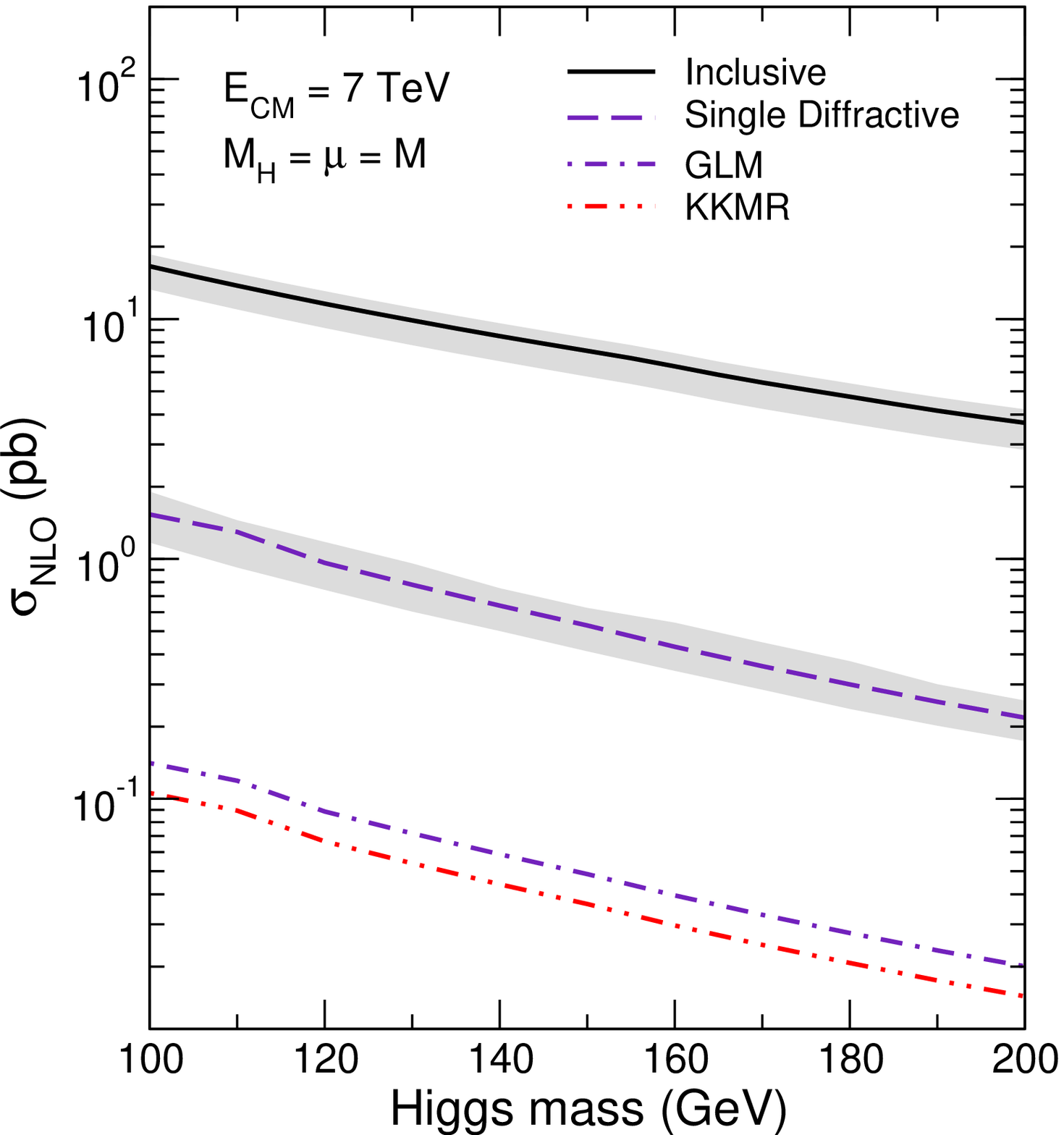}
\caption{\label{fig_7} The same as Fig.\ref{fig_lhc} for $\sqrt{s}$ = 7 TeV.}
\end{figure}

\begin{figure}
\includegraphics[width=\textwidth]{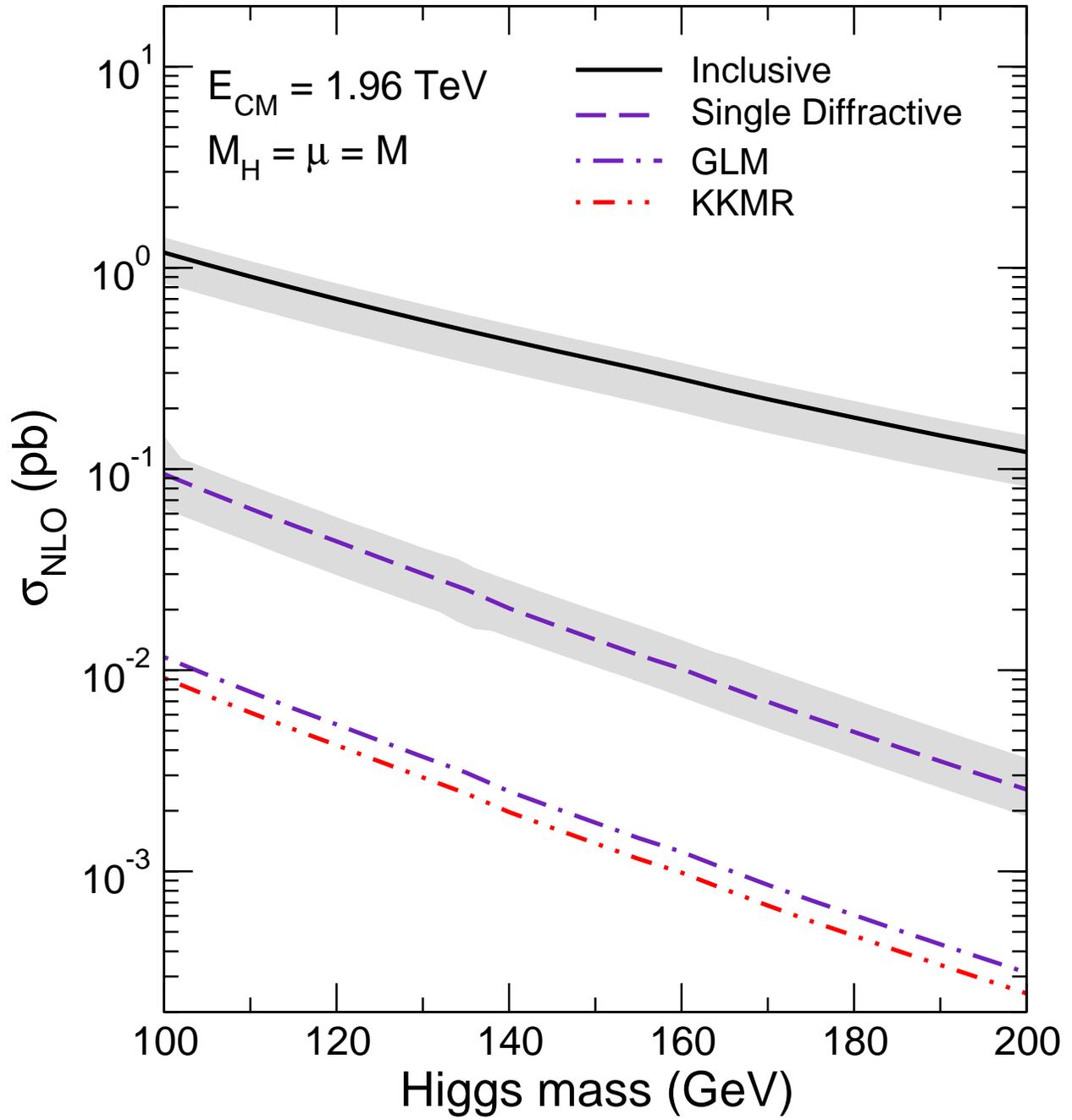}
\caption{\label{fig_tev} The same as Fig.\ref{fig_lhc} for the Tevatron energy.}
\end{figure}

Finally, also included are the electroweak two-loop corrections \cite{Actis:2008ts,*Actis:2008ug,*Actis:2008uh}, which enhance the total cross section by 5\% in comparison to the NNLO QCD cross section. In this way, the total cross section is computed with the addition of the EW corrections by
\begin{eqnarray}
\sigma_{\textrm{NLO}} \equiv \sigma_{\textrm{QCD+EW}} = \sigma_{\textrm{QCD}}(1 + \delta_{\textrm{EW}}).
\label{ew-corr}
\end{eqnarray}
The total cross sections for the inclusive process are shown by the solid curves in the Figs. \ref{fig_lhc}-\ref{fig_tev} for different collider energies. The gray bands around these curves express the variation of the renormalization and the factorization scales in the range 0.5$M_{H} < (\mu=M) < 4.0M_{H}$. Looking particularly to the results for the LHC, our results reproduce the values obtained in Ref.\cite{Spira:1995rr,*Dittmaier:2011ti}, although it is not the case for Ref.\cite{Erhan:2003za}\footnote{Comparing the results obtained in the Ref.\cite{Spira:1995rr,*Dittmaier:2011ti} and the curve presented in the Fig.6 in Ref.\cite{Erhan:2003za} for the total cross section in inclusive process, one can see that there is a disagreement between the results at $\sqrt{s}$ = 14 TeV, since the NNLO cross section for $M_{H}$ = 200 GeV in Ref.\cite{Erhan:2003za} is clearly smaller than that predicted in Ref.\cite{Spira:1995rr,Dittmaier:2011ti} at NLO.}.

\section{Diffractive production}
\label{sd-dpe}

For the diffractive process, the calculations are based on the IS formalism for diffractive hard scattering \cite{Ingelman:1984ns}. In this case, the Pomeron structure is taken into account by its quark and gluon content through the parametrization of the DPDF. The SD cross section is assumed to factorize into the Pomeron-hadron cross section and the Pomeron flux factor. In other words, it consists of three steps: first, a hard Pomeron is emitted from one of the protons in a small momentum transfer $|t|$, being this hadron detected in the final state; then, the second hadron scatters off the emitted Pomeron; during the Pomeron-hadron interaction, partons from the Pomeron interact with partons of the hadron, producing the Higgs boson. Accordingly, we will take into account absorptive effects, multiplying the diffractive cross section by a specific survival factor for each collider energy. The luminosity for the SD process reads
\begin{eqnarray}\nonumber
\frac{\dif{\cal{L}}_{\textrm{SD}}^{gi}}{\dif\tau} &=& \int^{1}_{\tau} \frac{\dif x}{x} \int \frac{\dif x_{\pom}}{x_{\pom}} F_{i/\pom/p}\left(x_{\pom},\frac{x}{x_{\pom}},M^{2}\right) g(\tau/x,M^{2}) \\
&+& \int^{1}_{\tau} \frac{\dif x}{x} \int \frac{\dif x_{\pom}}{x_{\pom}} g(x,M^{2}) F_{i/\pom/p}\left(x_{\pom},\frac{\tau}{xx_{\pom}},M^{2}\right),
\label{sdexp}
\end{eqnarray}
The Pomeron structure function $F_{i/\pom/p}$ is expressed by
\begin{eqnarray}
F_{i/\pom/ p} = f_{\pom/ p}(x_{\pom})f_{i/\pom}\left (\frac{x}{x_{\pom}},M^{2} \right ),
\label{func_pom}
\end{eqnarray}
with $f_{\pom/ p}(x_{\pom})$ being the Pomeron flux, and $f_{i/\pom} (\beta,\mu^{2})$ the parton distribution into the Pomeron, where $i$ stands for $g$, $q$, and $\bar{q}$.

In the estimates for the cross sections in Eq.(\ref{sdexp}), we consider a standard Pomeron flux from Regge phenomenology, which is constrained from the experimental analysis of the diffractive structure function. In this case, we apply the flux obtained with the H1 parametrization \cite{Aktas:2006hy}. The Pomeron structure function has been modeled in terms of a light flavor singlet distribution $\Sigma(x)$, i.e., the $u$, $d$ and $s$ quarks with their respective antiquarks. Also, it has a gluon distribution $g(z)$, with $z$ being the longitudinal momentum fraction of the parton in the hard subprocess. The gluon density is a constant at the starting evolution scale $Q^{2}_{0} = 2.5$ GeV$^{2}$. In our numerical calculations, we apply the cut $x < x_{\pom} \leq 0.05$ in agreement with the H1 parametrization. The Pomeron trajectory is assumed to be linear, $\alpha_{\pom}(t) = \alpha_{\pom}(0) + \alpha^{\prime}_{\pom}t$, with $\alpha^{\prime}_{\pom}$ and their uncertainties obtained from fits to H1 forward proton spectrometer (FPS) data \cite{Aktas:2006hx}. We choose $x_{\pom} \int^{t_{min}}_{t_{cut}} f_{\pom/p}\dif t = 1$ at $x_{\pom} = 0.003$, where $|t_{min}| \approx m^{2}_{p}x^{2}_{\pom}/(1 - x_{\pom})$ is the minimum kinematically accessible value of $|t|$, $m_{p}$ is the proton mass, and $|t_{cut}| = 1.0$ GeV$^{2}$ is the limit of the measurement. The H1 parametrization provides two different inputs for the fit of the partonic structure functions. As our curves show very close results using both fits, we chose the fit A to perform our predictions in this work.

\section{Gap Survival Probability}
\label{gsp}

In the diffractive cross sections [Eq.(\ref{sdexp})], we are further including the GSP $<\!|S|^{2}\!>$, being described in terms of absorptive corrections \cite{Bjorken:1991xr,Bjorken:1992er}. It can be estimated using the equation
\begin{eqnarray}
<\!|S|^2\!> = \frac{\int|{\cal{A}}\,(s,b)|^2\,e^{-\Omega (s,b)}\,\dif^{\,2}\!\boldsymbol{b}}{\int|{\cal{A}}\,(s,b)|^2\,\dif^{\,2}\!\boldsymbol{b}} ,
\end{eqnarray}
where $\cal{A}$ is the amplitude of the particular process of interest at the center-of-mass energy squared $s$ described in the impact parameter space $b$. The quantity $\Omega$ is the opacity (or optical density) of the interaction of the incoming hadrons. This suppression factor of a hard process accompanied by a rapidity gap does not depend only on the probability of the initial state survival, but it is also sensitive to the spatial distribution of partons inside the incoming hadrons, i.e., on the dynamics of the whole diffractive part of the scattering matrix. 

There are distinct approaches in the literature to compute the value of the $<\!|S|^2\!>$, predicting different probabilities for the diffractive Higgs boson production. Applying a survival factor to diffractive processes brings an uncertainty to the predictions for the production cross sections \cite{GayDucati:2007ps}, since there is no accurate prediction for the GSP, resulting in an imprecise predictions. Hence, we compare two different models for the GSP, being the most applied in other works, in order to investigate the available calculations of the survival factor to drive our predictions, and certainly the ones that will be studied to describe the future data. The first one is that of Refs. \cite{Khoze:2000vr,*Kaidalov:2001iz} (labeled KKMR), which considers a two-channel eikonal model that embodies pion-loop insertions in the Pomeron trajectory, diffractive dissociation and rescattering effects. Then, the survival probability is computed for single, central and double diffractive processes at different collider energies, assuming that the spatial distribution in impact parameter space is driven by the slope $B$ of the Pomeron-proton vertex. We will consider the value $<\!|S|^2\!>_{\mathrm{KKMR}}^{\mathrm{SD}}$ = 6\% (10\%) for the SD process in the LHC (Tevatron).

The second estimation for the survival factor is the model presented in Ref. \cite{Gotsman:1999xq,*Gotsman:2005rt} (labeled GLM), with a calculation for an eikonal single-channel approach. We take the case where the soft input is obtained directly from the measured values of $\sigma_{tot}$, $\sigma_{el}$ and hard radius $R_{H}$. The F1C approach was chosen to perform our predictions, resulting in a probability of $<\!|S|^2\!>^{\mathrm{SD}}_{\mathrm{GLM}}$ = 8.1\% (12.6\%) for the LHC (Tevatron) energy. We quote Ref. \cite{Gotsman:2005rt} for a detailed comparison between this approach and the Kaidalov-Khoze-Martin-Ryskin (KKMR) one, including further discussions on model dependence of inputs and consideration of multichannel calculations.

Unfortunately, these models only account for the GSP in the kinematical regime of the Tevatron or the LHC energies, i.e., for $\sqrt{s} = 1.8$ TeV and $14$ TeV. In order to make precise estimations with reliable values for the GSP, we chose to adopt a similar way to estimate the survival factor, in \% for the desired energy, following the approach of Ref.\cite{Machado:2007fr}
\begin{eqnarray}
<\!|S|^{2}\!>(\%) = \frac{a}{b + \ln\sqrt{s}},
\label{func-form}
\end{eqnarray}
with the parameters $a$ = 46.52 (30.77) and $b$ = -3.80 (-4.41) for the GLM (KKMR) model. Then, the Table \ref{tab-gsp} summarizes all the survival factors applied in the predictions for the SD process.

These particular models were chosen in order to indicate the uncertainty (model dependence) of the soft interaction effects. It is worth to mention that some implementations of GLM model include the results of a two- or three-channel calculation for $<\!|S|^2\!>$, which are considerably smaller than the one-channel approach \cite{Gotsman:2005rt}.

\section{Results and comments}
\label{results}

In this work we are mainly interested in the analysis of the cross sections for the SD Higgs boson production for different collider energies, bringing a rapidity gap in the final state as its main signature. Furthermore, the diffractive production can be an alternative way to detect the Higgs boson in hadron colliders, since it provides a higher signal-to-background ratio \cite{Khoze:2006uj}. Moreover, the SD production cross sections are presented in Figs. \ref{fig_lhc}-\ref{fig_tev}, being the results presented with no survival factor by the dashed curves, and including the GLM (dot-dashed) and KKMR models (double-dot-dashed). Additionally, some values of the production cross section are presented in Table \ref{tab1} for selected Higgs masses, showing specifically the values for the cross section with the adopted survival probabilities. As one can see, the production cross section in the kinematical regime of the Tevatron is very small, as expected. However, for higher energies the cross section reaches values of the order of 100 fb, showing that it may be possible to detect the Higgs boson in the LHC through the SD process. Besides, there are some detectors to be set up at the LHC experiments, and they will make it possible to tag the outgoing proton \cite{Albrow:2008pn,*Bonnet:2007pw,*Roland:2010ch} or to detect forward showers \cite{Albrow:2008az,*Lamsa:2009ej}. Then, the SD events can be an effective way to look for the Higgs boson at the LHC.

Still, to have a clear analysis of these results and to estimate the fraction of diffractive events, we compute the diffractive ratio for the Higgs boson mass of $M_{H}$ = 120 GeV, being presented in Table \ref{tab2}. As expected, the diffractive ratios are small and growing with the collider energy. Specifically, the GLM model shows a ratio nearly constant from the Tevatron energy until 7 TeV, and then growing for higher energies. However, the KKMR model shows a different behavior, presenting a decrease in the same region, and growing slowly at higher energies. This effect is observed due to our assumption for the survival factor for the collider energy of 7 and 8 TeV. However, a proper calculation of the survival factor for these energies may show a higher GSP than those presented in the Table \ref{tab-gsp}, which will increase the ratio, reaching a similar behavior as the result for the Tevatron. 

In fact, these values show that the SD events will have a very small rate in the LHC kinematical regime for a luminosity of a few fb$^{-1}$. It encourages the implementation of specific detectors in order to detect the rapidity gaps or the forward protons, since the Higgs boson discovery from its decay products is going to be more difficult in the inclusive production due to the huge background signal \cite{Rainwater:2002hm}.

Finally, in Figs. 5 and 6 we present the rapidity distribution of the Higgs boson for different collider energies. For higher energies, the distributions are clearly central, which shows that the contributions from the parton distribution function and the DPDF have larger values in central rapidity. Particularly, as the momentum fractions of the parton into the hadron A increases, the one of the parton into the hadron B (or the Pomeron) decreases uniformly, achieving a higher combined contribution for $y_{H} = 0$. However, this same behavior does not occur in the Tevatron kinematical regime, showing two distinct peaks in the distribution where the parton distribution function and DPDF have its higher combined contribution. In the results for mid energies (7 and 8 TeV), one can see that the distributions are still central, however showing very small peaks around $|y_{H}|$ = 2.

\section{Conclusions}
\label{concl}

In summary, we have evaluated estimations for the SD Higgs boson production in $pp$ collisions at the Tevatron and the LHC, considering the IS formalism with the introduction of rescattering corrections. We are using the Regge factorization to calculate the SD cross sections at NLO accuracy (QCD+EW). In particular, we take a parametrization from H1 Collaboration for the Pomeron structure function, extracted from their measurements of $F^{D(3)}_{2}$, with the results directly dependent on the quark singlet and gluon content of the Pomeron. For the available fits in this parametrization, we chose the fit A to perform our predictions. For instance, the cross sections are of about $\sigma_{\mathrm{SD}}$ = 0.4 (0.1) pb for $M_{H}$ = 120 GeV for $\sqrt{s}$ = 14 TeV (7 TeV). These cross sections are higher than that obtained from the $\gamma\gamma$ production mechanism, predicting a production cross section of 0.12 -- 0.18 fb \cite{Khoze:2001xm,*d'Enterria:2009er,*Miller:2007pc}, and even for the exclusive Higgs boson production \cite{Khoze:1997dr,*GayDucati:2010xi}. Moreover, comparing our estimations with no survival factor, we predict a cross section for $\sqrt{s}$ = 14 TeV higher than the previous results for the SD process \cite{Erhan:2003za}. In addition, the two models considered for the GSP have computed a survival factor that has a variation of 25\%. This difference is significative in order to perform reliable predictions for the Higgs boson production; however, it is expected that they are going to be tuned with the future data from the LHC experiments. In any case, our predictions with different survival factors may give a good estimation for the production cross section for the presented collider energies.

The SD production of the Higgs boson may not provide significative advantage in comparison to the inclusive production, since the background can not be suppressed using the same statements as in the double Pomeron exchange case. Nevertheless, the hadronic activity in the final state will be reduced in the SD events, increasing the possibility of observing the Higgs boson. Furthermore, the rapidity gaps may be observed in the LHC with the use of specific detectors, and then it can bring new data to be compared to the SD estimations. These results are the first NLO predictions for the single diffractive Higgs boson production, applying the GSP in the diffractive factorization. Besides, we have feasible values for the diffractive cross sections, and diffractive ratios as well, but the difference in the predictions is a bit high, which reveals that a study of the GSP for the multiple-Pomeron interactions in SD events is highly necessary. Therefore, we have presented updated estimations for the diffractive Higgs boson production, allowing the possibility to compare them to the future LHC data.

\begin{acknowledgments}
This work was supported by CNPq and FAPERGS, Brazil. We want to thanks M. V. T. Machado for useful comments. GGS would like to thank the Center for Particle Physics and Phenomenology (CP3) at Universit\'e catholique de Louvain for the hospitality.
\end{acknowledgments}

\pagebreak

\begin{table}
\begin{tabular}{ccc}
\toprule[0.07em]
$\sqrt{s}$ (TeV) & \multicolumn{2}{c}{$<\!|S|^{2}\!>$ (\%)} \tabularnewline
\midrule[0.07em]
                 & GLM  & KKMR                              \tabularnewline
\midrule[0.07em] 
1.96             & 12.3 & 9.7                               \tabularnewline
7.               & 9.2  & 6.9                               \tabularnewline
8.               & 8.9  & 6.7                               \tabularnewline
14.              & 8.1  & 6.0                               \tabularnewline
\bottomrule[0.07em]
\end{tabular}
\caption{\label{tab-gsp} Estimations for the survival factor for different collider energies obtained with Eq.(\ref{func-form}).}
\end{table}

\begin{table}
\begin{tabular}{cllll}
\toprule[0.07em]
Mass (GeV) & \multicolumn{4}{c}{$\sqrt{s}$ (TeV)}                       \tabularnewline
\midrule[0.07em]
           & 1.96       & 7.           & 8.            & 14.            \tabularnewline
\midrule[0.07em] 
120        & 5.36(4.23) & 88.59(66.44) & 119.70(90.11) & 346.43(256.62) \tabularnewline
140        & 2.57(2.02) & 58.69(44.02) & 81.43(61.30)  & 248.75(184.26) \tabularnewline
160        & 1.24(0.98) & 39.56(29.67) & 56.07(42.21)  & 183.06(135.60) \tabularnewline
180        & 0.60(0.47) & 27.60(20.70) & 40.23(30.28)  & 134.46(99.60)  \tabularnewline
200        & 0.31(0.24) & 19.96(14.97) & 29.10(21.90)  & 104.65(77.52)  \tabularnewline
\bottomrule[0.07em]
\end{tabular}
\caption{\label{tab1} Estimations for the production cross section (fb) at NLO accuracy for selected Higgs masses. The values are shown for the GLM (KKMR) models for the GSP and $M_{H} = \mu = M$.}
\end{table}

\begin{table}
\begin{tabular}{ccccc}
\toprule[0.07em]
Ratio (\%)          & \multicolumn{4}{c}{$\sqrt{s}$ (TeV)} \tabularnewline
\midrule[0.07em]
                    & 1.96 & 7.   & 8.   & 14.             \tabularnewline
\midrule[0.07em] 
$R_{\textrm{SD}}$   & 6.23 & 8.31 & 9.10 & 11.21           \tabularnewline
$R_{\textrm{GLM}}$  & 0.76 & 0.76 & 0.81 & 0.90            \tabularnewline
$R_{\textrm{KKMR}}$ & 0.60 & 0.57 & 0.61 & 0.67            \tabularnewline
\bottomrule[0.07em]
\end{tabular}
\caption{\label{tab2} Estimations for the diffractive ratios in different collider energies for a Higgs boson of $M_{H}$ = 120 GeV. The ratio is predicted for single diffractive events ($R_{\textrm{SD}}$) and for both models of the GSP ($R_{\textrm{GLM}}$ and $R_{\textrm{KKMR}}$) with $M_{H} = \mu = M$.}
\end{table}

\begin{figure}
\includegraphics[width=\textwidth]{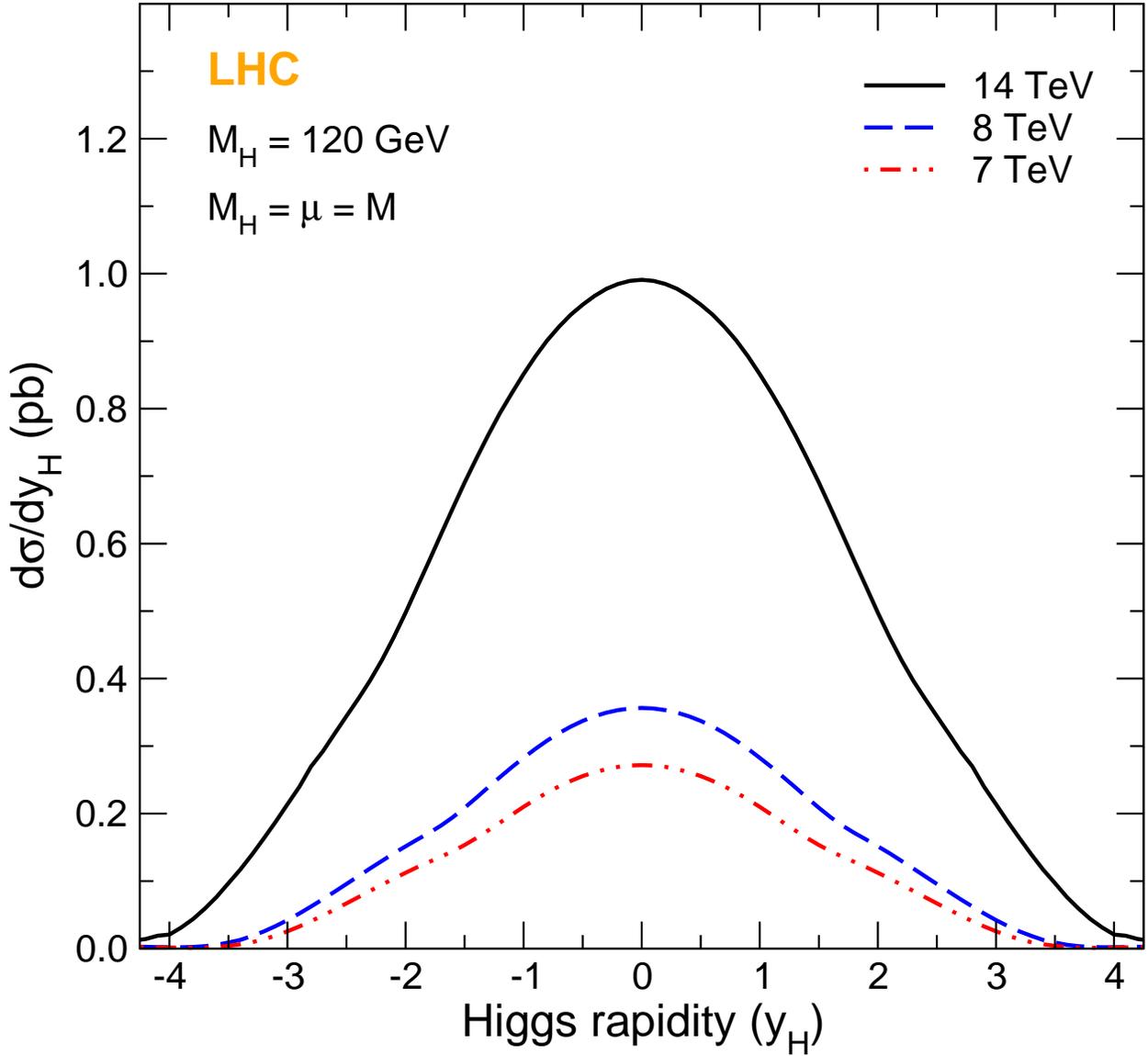}
\caption{\label{fig_yh_lhc} Rapidity distribution (pb) in single diffractive process of the Higgs boson for the current energy of the LHC (7 TeV) and for its future kinematical regimes.}
\end{figure}

\begin{figure}
\includegraphics[width=\textwidth]{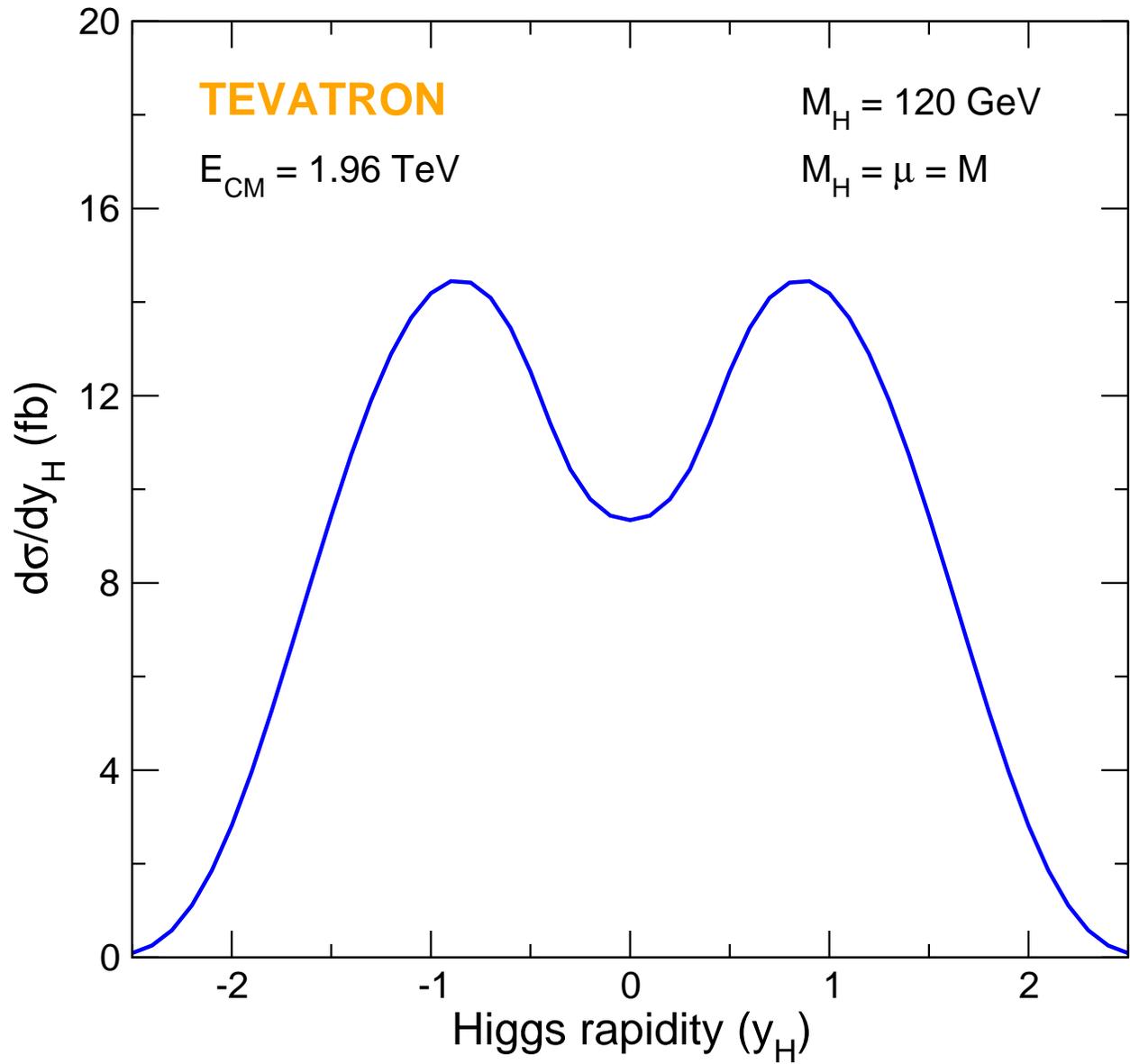}
\caption{\label{fig_yh_tev} Rapidity distribution (fb) in single diffractive process of the Higgs boson for the energy of the Tevatron (1.96 TeV).}
\end{figure}


\begin{thebibliography}{47}%
\makeatletter
\providecommand \@ifxundefined [1]{%
 \@ifx{#1\undefined}
}%
\providecommand \@ifnum [1]{%
 \ifnum #1\expandafter \@firstoftwo
 \else \expandafter \@secondoftwo
 \fi
}%
\providecommand \@ifx [1]{%
 \ifx #1\expandafter \@firstoftwo
 \else \expandafter \@secondoftwo
 \fi
}%
\providecommand \natexlab [1]{#1}%
\providecommand \enquote  [1]{``#1''}%
\providecommand \bibnamefont  [1]{#1}%
\providecommand \bibfnamefont [1]{#1}%
\providecommand \citenamefont [1]{#1}%
\providecommand \href@noop [0]{\@secondoftwo}%
\providecommand \href [0]{\begingroup \@sanitize@url \@href}%
\providecommand \@href[1]{\@@startlink{#1}\@@href}%
\providecommand \@@href[1]{\endgroup#1\@@endlink}%
\providecommand \@sanitize@url [0]{\catcode `\\12\catcode `\$12\catcode
  `\&12\catcode `\#12\catcode `\^12\catcode `\_12\catcode `\%12\relax}%
\providecommand \@@startlink[1]{}%
\providecommand \@@endlink[0]{}%
\providecommand \url  [0]{\begingroup\@sanitize@url \@url }%
\providecommand \@url [1]{\endgroup\@href {#1}{\urlprefix }}%
\providecommand \urlprefix  [0]{URL }%
\providecommand \Eprint [0]{\href }%
\providecommand \doibase [0]{http://dx.doi.org/}%
\providecommand \selectlanguage [0]{\@gobble}%
\providecommand \bibinfo  [0]{\@secondoftwo}%
\providecommand \bibfield  [0]{\@secondoftwo}%
\providecommand \translation [1]{[#1]}%
\providecommand \BibitemOpen [0]{}%
\providecommand \bibitemStop [0]{}%
\providecommand \bibitemNoStop [0]{.\EOS\space}%
\providecommand \EOS [0]{\spacefactor3000\relax}%
\providecommand \BibitemShut  [1]{\csname bibitem#1\endcsname}%
\let\auto@bib@innerbib\@empty
\bibitem [{\citenamefont {CDF}\ and\ \citenamefont
  {Collaborations}(2010)}]{:2010ar}%
  \BibitemOpen
  \bibfield  {author} {\bibinfo {author} {\bibnamefont {CDF}}\ and\ \bibinfo
  {author} {\bibfnamefont {D.}~\bibnamefont {Collaborations}},\ }\href@noop {}
  {\enquote {\bibinfo {title} {{Combined CDF and D0 Upper Limits on Standard
  Model Higgs- Boson Production with up to 6.7 fb$^{-1}$ of Data}},}\ }
  (\bibinfo {year} {2010}),\ \Eprint {http://arxiv.org/abs/1007.4587}
  {arXiv:1007.4587 [hep-ex]} \BibitemShut {NoStop}%
\bibitem [{\citenamefont {Tsukerman}(2010)}]{Collaboration:2010dk}%
  \BibitemOpen
  \bibfield  {author} {\bibinfo {author} {\bibfnamefont {I.}~\bibnamefont
  {Tsukerman}},\ }\href@noop {} {\enquote {\bibinfo {title} {{SM Higgs boson
  searches in the early ATLAS data}},}\ } (\bibinfo {year} {2010}),\ \Eprint
  {http://arxiv.org/abs/1012.0694} {arXiv:1012.0694 [hep-ex]} \BibitemShut
  {NoStop}%
\bibitem [{\citenamefont {Collins}(1977)}]{Collins:1977jy}%
  \BibitemOpen
  \bibfield  {author} {\bibinfo {author} {\bibfnamefont {P.~D.~B.}\
  \bibnamefont {Collins}},\ }\href@noop {} {\emph {\bibinfo {title} {{An
  Introduction to Regge Theory and High-Energy Physics}}}}\ (\bibinfo
  {publisher} {Cambridge University Press},\ \bibinfo {year}
  {1977})\BibitemShut {NoStop}%
\bibitem [{\citenamefont {Foldy}\ and\ \citenamefont
  {Peierls}(1963)}]{Foldy:1963zz}%
  \BibitemOpen
  \bibfield  {author} {\bibinfo {author} {\bibfnamefont {L.~L.}\ \bibnamefont
  {Foldy}}\ and\ \bibinfo {author} {\bibfnamefont {R.~F.}\ \bibnamefont
  {Peierls}},\ }\href {\doibase 10.1103/PhysRev.130.1585} {\bibfield  {journal}
  {\bibinfo  {journal} {Phys. Rev.}\ }\textbf {\bibinfo {volume} {130}},\
  \bibinfo {pages} {1585} (\bibinfo {year} {1963})}\BibitemShut {NoStop}%
\bibitem [{\citenamefont {Donnachie}\ and\ \citenamefont
  {Landshoff}(1992)}]{Donnachie:1992ny}%
  \BibitemOpen
  \bibfield  {author} {\bibinfo {author} {\bibfnamefont {A.}~\bibnamefont
  {Donnachie}}\ and\ \bibinfo {author} {\bibfnamefont {P.~V.}\ \bibnamefont
  {Landshoff}},\ }\href {\doibase 10.1016/0370-2693(92)90832-O} {\bibfield
  {journal} {\bibinfo  {journal} {Phys. Lett.}\ }\textbf {\bibinfo {volume}
  {B296}},\ \bibinfo {pages} {227} (\bibinfo {year} {1992})},\ \Eprint
  {http://arxiv.org/abs/hep-ph/9209205} {arXiv:hep-ph/9209205} \BibitemShut
  {NoStop}%
\bibitem [{\citenamefont {Ingelman}\ and\ \citenamefont
  {Schlein}(1985)}]{Ingelman:1984ns}%
  \BibitemOpen
  \bibfield  {author} {\bibinfo {author} {\bibfnamefont {G.}~\bibnamefont
  {Ingelman}}\ and\ \bibinfo {author} {\bibfnamefont {P.~E.}\ \bibnamefont
  {Schlein}},\ }\href {\doibase 10.1016/0370-2693(85)91181-5} {\bibfield
  {journal} {\bibinfo  {journal} {Phys. Lett.}\ }\textbf {\bibinfo {volume}
  {B152}},\ \bibinfo {pages} {256} (\bibinfo {year} {1985})}\BibitemShut
  {NoStop}%
\bibitem [{\citenamefont {Aktas}\ \emph
  {et~al.}(2006{\natexlab{a}})\citenamefont {Aktas} \emph
  {et~al.}}]{Aktas:2006hy}%
  \BibitemOpen
  \bibfield  {author} {\bibinfo {author} {\bibfnamefont {A.}~\bibnamefont
  {Aktas}} \emph {et~al.} (\bibinfo {collaboration} {H1}),\ }\href {\doibase
  10.1140/epjc/s10052-006-0035-3} {\bibfield  {journal} {\bibinfo  {journal}
  {Eur. Phys. J.}\ }\textbf {\bibinfo {volume} {C48}},\ \bibinfo {pages} {715}
  (\bibinfo {year} {2006}{\natexlab{a}})},\ \Eprint
  {http://arxiv.org/abs/hep-ex/0606004} {arXiv:hep-ex/0606004} \BibitemShut
  {NoStop}%
\bibitem [{\citenamefont {Gay~Ducati}\ \emph {et~al.}(2007)\citenamefont
  {Gay~Ducati}, \citenamefont {Machado},\ and\ \citenamefont
  {Machado}}]{GayDucati:2007ps}%
  \BibitemOpen
  \bibfield  {author} {\bibinfo {author} {\bibfnamefont {M.~B.}\ \bibnamefont
  {Gay~Ducati}}, \bibinfo {author} {\bibfnamefont {M.~M.}\ \bibnamefont
  {Machado}}, \ and\ \bibinfo {author} {\bibfnamefont {M.~V.~T.}\ \bibnamefont
  {Machado}},\ }\href {\doibase 10.1103/PhysRevD.75.114013} {\bibfield
  {journal} {\bibinfo  {journal} {Phys. Rev.}\ }\textbf {\bibinfo {volume}
  {D75}},\ \bibinfo {pages} {114013} (\bibinfo {year} {2007})},\ \Eprint
  {http://arxiv.org/abs/hep-ph/0703315} {arXiv:hep-ph/0703315} \BibitemShut
  {NoStop}%
\bibitem [{\citenamefont {Kopeliovich}\ \emph {et~al.}(2005)\citenamefont
  {Kopeliovich}, \citenamefont {Nemchik}, \citenamefont {Potashnikova},
  \citenamefont {Johnson},\ and\ \citenamefont {Schmidt}}]{Kopeliovich:2005ym}%
  \BibitemOpen
  \bibfield  {author} {\bibinfo {author} {\bibfnamefont {B.~Z.}\ \bibnamefont
  {Kopeliovich}}, \bibinfo {author} {\bibfnamefont {J.}~\bibnamefont
  {Nemchik}}, \bibinfo {author} {\bibfnamefont {I.~K.}\ \bibnamefont
  {Potashnikova}}, \bibinfo {author} {\bibfnamefont {M.~B.}\ \bibnamefont
  {Johnson}}, \ and\ \bibinfo {author} {\bibfnamefont {I.}~\bibnamefont
  {Schmidt}},\ }\href {\doibase 10.1103/PhysRevC.72.054606} {\bibfield
  {journal} {\bibinfo  {journal} {Phys. Rev.}\ }\textbf {\bibinfo {volume}
  {C72}},\ \bibinfo {pages} {054606} (\bibinfo {year} {2005})},\ \Eprint
  {http://arxiv.org/abs/hep-ph/0501260} {arXiv:hep-ph/0501260} \BibitemShut
  {NoStop}%
\bibitem [{\citenamefont {Chehime}\ \emph {et~al.}(1992)\citenamefont {Chehime}
  \emph {et~al.}}]{Chehime:1992bp}%
  \BibitemOpen
  \bibfield  {author} {\bibinfo {author} {\bibfnamefont {H.}~\bibnamefont
  {Chehime}} \emph {et~al.},\ }\href {\doibase 10.1016/0370-2693(92)91794-A}
  {\bibfield  {journal} {\bibinfo  {journal} {Phys. Lett.}\ }\textbf {\bibinfo
  {volume} {B286}},\ \bibinfo {pages} {397} (\bibinfo {year}
  {1992})}\BibitemShut {NoStop}%
\bibitem [{\citenamefont {Bjorken}(1992)}]{Bjorken:1991xr}%
  \BibitemOpen
  \bibfield  {author} {\bibinfo {author} {\bibfnamefont {J.~D.}\ \bibnamefont
  {Bjorken}},\ }\href {\doibase 10.1142/S0217751X92001885} {\bibfield
  {journal} {\bibinfo  {journal} {Int. J. Mod. Phys.}\ }\textbf {\bibinfo
  {volume} {A7}},\ \bibinfo {pages} {4189} (\bibinfo {year}
  {1992})}\BibitemShut {NoStop}%
\bibitem [{\citenamefont {Bjorken}(1993)}]{Bjorken:1992er}%
  \BibitemOpen
  \bibfield  {author} {\bibinfo {author} {\bibfnamefont {J.~D.}\ \bibnamefont
  {Bjorken}},\ }\href {\doibase 10.1103/PhysRevD.47.101} {\bibfield  {journal}
  {\bibinfo  {journal} {Phys. Rev.}\ }\textbf {\bibinfo {volume} {D47}},\
  \bibinfo {pages} {101} (\bibinfo {year} {1993})}\BibitemShut {NoStop}%
\bibitem [{\citenamefont {Gay~Ducati}\ \emph
  {et~al.}(2010{\natexlab{a}})\citenamefont {Gay~Ducati}, \citenamefont
  {Machado},\ and\ \citenamefont {Machado}}]{GayDucati:2010vu}%
  \BibitemOpen
  \bibfield  {author} {\bibinfo {author} {\bibfnamefont {M.~B.}\ \bibnamefont
  {Gay~Ducati}}, \bibinfo {author} {\bibfnamefont {M.~M.}\ \bibnamefont
  {Machado}}, \ and\ \bibinfo {author} {\bibfnamefont {M.~V.~T.}\ \bibnamefont
  {Machado}},\ }\href {\doibase 10.1103/PhysRevD.81.054034} {\bibfield
  {journal} {\bibinfo  {journal} {Phys. Rev.}\ }\textbf {\bibinfo {volume}
  {D81}},\ \bibinfo {pages} {054034} (\bibinfo {year} {2010}{\natexlab{a}})},\
  \Eprint {http://arxiv.org/abs/1002.4043} {arXiv:1002.4043 [hep-ph]}
  \BibitemShut {NoStop}%
\bibitem [{\citenamefont {Gay~Ducati}\ \emph
  {et~al.}(2010{\natexlab{b}})\citenamefont {Gay~Ducati}, \citenamefont
  {Machado},\ and\ \citenamefont {Machado}}]{GayDucati:2009rr}%
  \BibitemOpen
  \bibfield  {author} {\bibinfo {author} {\bibfnamefont {M.~B.}\ \bibnamefont
  {Gay~Ducati}}, \bibinfo {author} {\bibfnamefont {M.~M.}\ \bibnamefont
  {Machado}}, \ and\ \bibinfo {author} {\bibfnamefont {M.~V.~T.}\ \bibnamefont
  {Machado}},\ }\href {\doibase 10.1016/j.physletb.2009.12.025} {\bibfield
  {journal} {\bibinfo  {journal} {Phys. Lett.}\ }\textbf {\bibinfo {volume}
  {B683}},\ \bibinfo {pages} {150} (\bibinfo {year} {2010}{\natexlab{b}})},\
  \Eprint {http://arxiv.org/abs/0908.0507} {arXiv:0908.0507 [hep-ph]}
  \BibitemShut {NoStop}%
\bibitem [{\citenamefont {Carena}\ and\ \citenamefont
  {Haber}(2003)}]{Carena:2002es}%
  \BibitemOpen
  \bibfield  {author} {\bibinfo {author} {\bibfnamefont {M.~S.}\ \bibnamefont
  {Carena}}\ and\ \bibinfo {author} {\bibfnamefont {H.~E.}\ \bibnamefont
  {Haber}},\ }\href {\doibase 10.1016/S0146-6410(02)00177-1} {\bibfield
  {journal} {\bibinfo  {journal} {Prog. Part. Nucl. Phys.}\ }\textbf {\bibinfo
  {volume} {50}},\ \bibinfo {pages} {63} (\bibinfo {year} {2003})},\ \Eprint
  {http://arxiv.org/abs/hep-ph/0208209} {arXiv:hep-ph/0208209} \BibitemShut
  {NoStop}%
\bibitem [{\citenamefont {Hahn}\ \emph {et~al.}(2006)\citenamefont {Hahn},
  \citenamefont {Heinemeyer}, \citenamefont {Maltoni}, \citenamefont
  {Weiglein},\ and\ \citenamefont {Willenbrock}}]{Hahn:2006my}%
  \BibitemOpen
  \bibfield  {author} {\bibinfo {author} {\bibfnamefont {T.}~\bibnamefont
  {Hahn}}, \bibinfo {author} {\bibfnamefont {S.}~\bibnamefont {Heinemeyer}},
  \bibinfo {author} {\bibfnamefont {F.}~\bibnamefont {Maltoni}}, \bibinfo
  {author} {\bibfnamefont {G.}~\bibnamefont {Weiglein}}, \ and\ \bibinfo
  {author} {\bibfnamefont {S.}~\bibnamefont {Willenbrock}},\ }\href@noop {}
  {\enquote {\bibinfo {title} {{SM and MSSM Higgs boson production cross
  sections at the Tevatron and the LHC}},}\ } (\bibinfo {year} {2006}),\
  \Eprint {http://arxiv.org/abs/hep-ph/0607308} {arXiv:hep-ph/0607308}
  \BibitemShut {NoStop}%
\bibitem [{\citenamefont {Duperrin}(2009)}]{Duperrin:2008in}%
  \BibitemOpen
  \bibfield  {author} {\bibinfo {author} {\bibfnamefont {A.}~\bibnamefont
  {Duperrin}},\ }\href {\doibase 10.1140/epjc/s10052-008-0719-y} {\bibfield
  {journal} {\bibinfo  {journal} {Eur. Phys. J.}\ }\textbf {\bibinfo {volume}
  {C59}},\ \bibinfo {pages} {297} (\bibinfo {year} {2009})},\ \Eprint
  {http://arxiv.org/abs/0805.3624} {arXiv:0805.3624 [hep-ex]} \BibitemShut
  {NoStop}%
\bibitem [{\citenamefont {Spira}\ \emph {et~al.}(1995)\citenamefont {Spira},
  \citenamefont {Djouadi}, \citenamefont {Graudenz},\ and\ \citenamefont
  {Zerwas}}]{Spira:1995rr}%
  \BibitemOpen
  \bibfield  {author} {\bibinfo {author} {\bibfnamefont {M.}~\bibnamefont
  {Spira}}, \bibinfo {author} {\bibfnamefont {A.}~\bibnamefont {Djouadi}},
  \bibinfo {author} {\bibfnamefont {D.}~\bibnamefont {Graudenz}}, \ and\
  \bibinfo {author} {\bibfnamefont {P.~M.}\ \bibnamefont {Zerwas}},\ }\href
  {\doibase 10.1016/0550-3213(95)00379-7} {\bibfield  {journal} {\bibinfo
  {journal} {Nucl. Phys.}\ }\textbf {\bibinfo {volume} {B453}},\ \bibinfo
  {pages} {17} (\bibinfo {year} {1995})},\ \Eprint
  {http://arxiv.org/abs/hep-ph/9504378} {arXiv:hep-ph/9504378} \BibitemShut
  {NoStop}%
\bibitem [{\citenamefont {Martin}\ \emph
  {et~al.}(2009{\natexlab{a}})\citenamefont {Martin}, \citenamefont {Stirling},
  \citenamefont {Thorne},\ and\ \citenamefont {Watt}}]{Martin:2009bu}%
  \BibitemOpen
  \bibfield  {author} {\bibinfo {author} {\bibfnamefont {A.~D.}\ \bibnamefont
  {Martin}}, \bibinfo {author} {\bibfnamefont {W.~J.}\ \bibnamefont
  {Stirling}}, \bibinfo {author} {\bibfnamefont {R.~S.}\ \bibnamefont
  {Thorne}}, \ and\ \bibinfo {author} {\bibfnamefont {G.}~\bibnamefont
  {Watt}},\ }\href {\doibase 10.1140/epjc/s10052-009-1164-2} {\bibfield
  {journal} {\bibinfo  {journal} {Eur. Phys. J.}\ }\textbf {\bibinfo {volume}
  {C64}},\ \bibinfo {pages} {653} (\bibinfo {year} {2009}{\natexlab{a}})},\
  \Eprint {http://arxiv.org/abs/0905.3531} {arXiv:0905.3531 [hep-ph]}
  \BibitemShut {NoStop}%
\bibitem [{\citenamefont {Martin}\ \emph
  {et~al.}(2009{\natexlab{b}})\citenamefont {Martin}, \citenamefont {Stirling},
  \citenamefont {Thorne},\ and\ \citenamefont {Watt}}]{Martin:2009iq}%
  \BibitemOpen
  \bibfield  {author} {\bibinfo {author} {\bibfnamefont {A.~D.}\ \bibnamefont
  {Martin}}, \bibinfo {author} {\bibfnamefont {W.~J.}\ \bibnamefont
  {Stirling}}, \bibinfo {author} {\bibfnamefont {R.~S.}\ \bibnamefont
  {Thorne}}, \ and\ \bibinfo {author} {\bibfnamefont {G.}~\bibnamefont
  {Watt}},\ }\href {\doibase 10.1140/epjc/s10052-009-1072-5} {\bibfield
  {journal} {\bibinfo  {journal} {Eur. Phys. J.}\ }\textbf {\bibinfo {volume}
  {C63}},\ \bibinfo {pages} {189} (\bibinfo {year} {2009}{\natexlab{b}})},\
  \Eprint {http://arxiv.org/abs/0901.0002} {arXiv:0901.0002 [hep-ph]}
  \BibitemShut {NoStop}%
\bibitem [{\citenamefont {Dawson}(1991)}]{Dawson:1990zj}%
  \BibitemOpen
  \bibfield  {author} {\bibinfo {author} {\bibfnamefont {S.}~\bibnamefont
  {Dawson}},\ }\href {\doibase 10.1016/0550-3213(91)90061-2} {\bibfield
  {journal} {\bibinfo  {journal} {Nucl. Phys.}\ }\textbf {\bibinfo {volume}
  {B359}},\ \bibinfo {pages} {283} (\bibinfo {year} {1991})}\BibitemShut
  {NoStop}%
\bibitem [{\citenamefont {Gluck}\ \emph {et~al.}(1998)\citenamefont {Gluck},
  \citenamefont {Reya},\ and\ \citenamefont {Vogt}}]{Gluck:1998xa}%
  \BibitemOpen
  \bibfield  {author} {\bibinfo {author} {\bibfnamefont {M.}~\bibnamefont
  {Gluck}}, \bibinfo {author} {\bibfnamefont {E.}~\bibnamefont {Reya}}, \ and\
  \bibinfo {author} {\bibfnamefont {A.}~\bibnamefont {Vogt}},\ }\href {\doibase
  10.1007/s100520050289} {\bibfield  {journal} {\bibinfo  {journal} {Eur. Phys.
  J.}\ }\textbf {\bibinfo {volume} {C5}},\ \bibinfo {pages} {461} (\bibinfo
  {year} {1998})},\ \Eprint {http://arxiv.org/abs/hep-ph/9806404}
  {arXiv:hep-ph/9806404} \BibitemShut {NoStop}%
\bibitem [{\citenamefont {Graudenz}\ \emph {et~al.}(1993)\citenamefont
  {Graudenz}, \citenamefont {Spira},\ and\ \citenamefont
  {Zerwas}}]{Graudenz:1992pv}%
  \BibitemOpen
  \bibfield  {author} {\bibinfo {author} {\bibfnamefont {D.}~\bibnamefont
  {Graudenz}}, \bibinfo {author} {\bibfnamefont {M.}~\bibnamefont {Spira}}, \
  and\ \bibinfo {author} {\bibfnamefont {P.~M.}\ \bibnamefont {Zerwas}},\
  }\href {\doibase 10.1103/PhysRevLett.70.1372} {\bibfield  {journal} {\bibinfo
   {journal} {Phys. Rev. Lett.}\ }\textbf {\bibinfo {volume} {70}},\ \bibinfo
  {pages} {1372} (\bibinfo {year} {1993})}\BibitemShut {NoStop}%
\bibitem [{\citenamefont {Altarelli}\ and\ \citenamefont
  {Parisi}(1977)}]{Altarelli:1977zs}%
  \BibitemOpen
  \bibfield  {author} {\bibinfo {author} {\bibfnamefont {G.}~\bibnamefont
  {Altarelli}}\ and\ \bibinfo {author} {\bibfnamefont {G.}~\bibnamefont
  {Parisi}},\ }\href {\doibase 10.1016/0550-3213(77)90384-4} {\bibfield
  {journal} {\bibinfo  {journal} {Nucl. Phys.}\ }\textbf {\bibinfo {volume}
  {B126}},\ \bibinfo {pages} {298} (\bibinfo {year} {1977})}\BibitemShut
  {NoStop}%
\bibitem [{\citenamefont {Actis}\ \emph {et~al.}(2009)\citenamefont {Actis},
  \citenamefont {Passarino}, \citenamefont {Sturm},\ and\ \citenamefont
  {Uccirati}}]{Actis:2008ts}%
  \BibitemOpen
  \bibfield  {author} {\bibinfo {author} {\bibfnamefont {S.}~\bibnamefont
  {Actis}}, \bibinfo {author} {\bibfnamefont {G.}~\bibnamefont {Passarino}},
  \bibinfo {author} {\bibfnamefont {C.}~\bibnamefont {Sturm}}, \ and\ \bibinfo
  {author} {\bibfnamefont {S.}~\bibnamefont {Uccirati}},\ }\href {\doibase
  10.1016/j.nuclphysb.2008.11.024} {\bibfield  {journal} {\bibinfo  {journal}
  {Nucl. Phys.}\ }\textbf {\bibinfo {volume} {B811}},\ \bibinfo {pages} {182}
  (\bibinfo {year} {2009})},\ \Eprint {http://arxiv.org/abs/0809.3667}
  {arXiv:0809.3667 [hep-ph]} \BibitemShut {NoStop}%
\bibitem [{\citenamefont {Actis}\ \emph
  {et~al.}(2008{\natexlab{a}})\citenamefont {Actis}, \citenamefont {Passarino},
  \citenamefont {Sturm},\ and\ \citenamefont {Uccirati}}]{Actis:2008ug}%
  \BibitemOpen
  \bibfield  {author} {\bibinfo {author} {\bibfnamefont {S.}~\bibnamefont
  {Actis}}, \bibinfo {author} {\bibfnamefont {G.}~\bibnamefont {Passarino}},
  \bibinfo {author} {\bibfnamefont {C.}~\bibnamefont {Sturm}}, \ and\ \bibinfo
  {author} {\bibfnamefont {S.}~\bibnamefont {Uccirati}},\ }\href {\doibase
  10.1016/j.physletb.2008.10.018} {\bibfield  {journal} {\bibinfo  {journal}
  {Phys. Lett.}\ }\textbf {\bibinfo {volume} {B670}},\ \bibinfo {pages} {12}
  (\bibinfo {year} {2008}{\natexlab{a}})},\ \Eprint
  {http://arxiv.org/abs/0809.1301} {arXiv:0809.1301 [hep-ph]} \BibitemShut
  {NoStop}%
\bibitem [{\citenamefont {Actis}\ \emph
  {et~al.}(2008{\natexlab{b}})\citenamefont {Actis}, \citenamefont {Passarino},
  \citenamefont {Sturm},\ and\ \citenamefont {Uccirati}}]{Actis:2008uh}%
  \BibitemOpen
  \bibfield  {author} {\bibinfo {author} {\bibfnamefont {S.}~\bibnamefont
  {Actis}}, \bibinfo {author} {\bibfnamefont {G.}~\bibnamefont {Passarino}},
  \bibinfo {author} {\bibfnamefont {C.}~\bibnamefont {Sturm}}, \ and\ \bibinfo
  {author} {\bibfnamefont {S.}~\bibnamefont {Uccirati}},\ }\href {\doibase
  10.1016/j.physletb.2008.09.028} {\bibfield  {journal} {\bibinfo  {journal}
  {Phys. Lett.}\ }\textbf {\bibinfo {volume} {B669}},\ \bibinfo {pages} {62}
  (\bibinfo {year} {2008}{\natexlab{b}})},\ \Eprint
  {http://arxiv.org/abs/0809.1302} {arXiv:0809.1302 [hep-ph]} \BibitemShut
  {NoStop}%
\bibitem [{\citenamefont {Dittmaier}\ \emph {et~al.}(2011)\citenamefont
  {Dittmaier} \emph {et~al.}}]{Dittmaier:2011ti}%
  \BibitemOpen
  \bibfield  {author} {\bibinfo {author} {\bibfnamefont {S.}~\bibnamefont
  {Dittmaier}} \emph {et~al.} (\bibinfo {collaboration} {LHC Higgs Cross
  Section Working Group}),\ }\href@noop {} {\enquote {\bibinfo {title}
  {{Handbook of LHC Higgs Cross Sections: 1. Inclusive Observables}},}\ }
  (\bibinfo {year} {2011}),\ \Eprint {http://arxiv.org/abs/1101.0593}
  {arXiv:1101.0593 [hep-ph]} \BibitemShut {NoStop}%
\bibitem [{\citenamefont {Erhan}\ \emph {et~al.}(2003)\citenamefont {Erhan},
  \citenamefont {Kim},\ and\ \citenamefont {Schlein}}]{Erhan:2003za}%
  \BibitemOpen
  \bibfield  {author} {\bibinfo {author} {\bibfnamefont {S.}~\bibnamefont
  {Erhan}}, \bibinfo {author} {\bibfnamefont {V.~T.}\ \bibnamefont {Kim}}, \
  and\ \bibinfo {author} {\bibfnamefont {P.~E.}\ \bibnamefont {Schlein}},\
  }\href@noop {} {\enquote {\bibinfo {title} {{Central Higgs production at LHC
  from single pomeron- exchange}},}\ } (\bibinfo {year} {2003}),\ \Eprint
  {http://arxiv.org/abs/hep-ph/0312342} {arXiv:hep-ph/0312342} \BibitemShut
  {NoStop}%
\bibitem [{\citenamefont {Aktas}\ \emph
  {et~al.}(2006{\natexlab{b}})\citenamefont {Aktas} \emph
  {et~al.}}]{Aktas:2006hx}%
  \BibitemOpen
  \bibfield  {author} {\bibinfo {author} {\bibfnamefont {A.}~\bibnamefont
  {Aktas}} \emph {et~al.} (\bibinfo {collaboration} {H1}),\ }\href {\doibase
  10.1140/epjc/s10052-006-0046-0} {\bibfield  {journal} {\bibinfo  {journal}
  {Eur. Phys. J.}\ }\textbf {\bibinfo {volume} {C48}},\ \bibinfo {pages} {749}
  (\bibinfo {year} {2006}{\natexlab{b}})},\ \Eprint
  {http://arxiv.org/abs/hep-ex/0606003} {arXiv:hep-ex/0606003} \BibitemShut
  {NoStop}%
\bibitem [{\citenamefont {Khoze}\ \emph {et~al.}(2001)\citenamefont {Khoze},
  \citenamefont {Martin},\ and\ \citenamefont {Ryskin}}]{Khoze:2000vr}%
  \BibitemOpen
  \bibfield  {author} {\bibinfo {author} {\bibfnamefont {V.~A.}\ \bibnamefont
  {Khoze}}, \bibinfo {author} {\bibfnamefont {A.~D.}\ \bibnamefont {Martin}}, \
  and\ \bibinfo {author} {\bibfnamefont {M.~G.}\ \bibnamefont {Ryskin}},\
  }\href@noop {} {\bibfield  {journal} {\bibinfo  {journal} {Nucl. Phys. Proc.
  Suppl.}\ }\textbf {\bibinfo {volume} {99B}},\ \bibinfo {pages} {213}
  (\bibinfo {year} {2001})},\ \Eprint {http://arxiv.org/abs/hep-ph/0011319}
  {arXiv:hep-ph/0011319} \BibitemShut {NoStop}%
\bibitem [{\citenamefont {Kaidalov}\ \emph {et~al.}(2001)\citenamefont
  {Kaidalov}, \citenamefont {Khoze}, \citenamefont {Martin},\ and\
  \citenamefont {Ryskin}}]{Kaidalov:2001iz}%
  \BibitemOpen
  \bibfield  {author} {\bibinfo {author} {\bibfnamefont {A.~B.}\ \bibnamefont
  {Kaidalov}}, \bibinfo {author} {\bibfnamefont {V.~A.}\ \bibnamefont {Khoze}},
  \bibinfo {author} {\bibfnamefont {A.~D.}\ \bibnamefont {Martin}}, \ and\
  \bibinfo {author} {\bibfnamefont {M.~G.}\ \bibnamefont {Ryskin}},\ }\href
  {\doibase 10.1007/s100520100751} {\bibfield  {journal} {\bibinfo  {journal}
  {Eur. Phys. J.}\ }\textbf {\bibinfo {volume} {C21}},\ \bibinfo {pages} {521}
  (\bibinfo {year} {2001})},\ \Eprint {http://arxiv.org/abs/hep-ph/0105145}
  {arXiv:hep-ph/0105145} \BibitemShut {NoStop}%
\bibitem [{\citenamefont {Gotsman}\ \emph {et~al.}(1999)\citenamefont
  {Gotsman}, \citenamefont {Levin},\ and\ \citenamefont
  {Maor}}]{Gotsman:1999xq}%
  \BibitemOpen
  \bibfield  {author} {\bibinfo {author} {\bibfnamefont {E.}~\bibnamefont
  {Gotsman}}, \bibinfo {author} {\bibfnamefont {E.}~\bibnamefont {Levin}}, \
  and\ \bibinfo {author} {\bibfnamefont {U.}~\bibnamefont {Maor}},\ }\href
  {\doibase 10.1103/PhysRevD.60.094011} {\bibfield  {journal} {\bibinfo
  {journal} {Phys. Rev.}\ }\textbf {\bibinfo {volume} {D60}},\ \bibinfo {pages}
  {094011} (\bibinfo {year} {1999})},\ \Eprint
  {http://arxiv.org/abs/hep-ph/9902294} {arXiv:hep-ph/9902294} \BibitemShut
  {NoStop}%
\bibitem [{\citenamefont {Gotsman}\ \emph {et~al.}(2005)\citenamefont
  {Gotsman}, \citenamefont {Levin}, \citenamefont {Maor}, \citenamefont
  {Naftali},\ and\ \citenamefont {Prygarin}}]{Gotsman:2005rt}%
  \BibitemOpen
  \bibfield  {author} {\bibinfo {author} {\bibfnamefont {E.}~\bibnamefont
  {Gotsman}}, \bibinfo {author} {\bibfnamefont {E.}~\bibnamefont {Levin}},
  \bibinfo {author} {\bibfnamefont {U.}~\bibnamefont {Maor}}, \bibinfo {author}
  {\bibfnamefont {E.}~\bibnamefont {Naftali}}, \ and\ \bibinfo {author}
  {\bibfnamefont {A.}~\bibnamefont {Prygarin}},\ }\href@noop {} {\enquote
  {\bibinfo {title} {{Survival probability of large rapidity gaps}},}\ }
  (\bibinfo {year} {2005}),\ \Eprint {http://arxiv.org/abs/hep-ph/0511060}
  {arXiv:hep-ph/0511060} \BibitemShut {NoStop}%
\bibitem [{\citenamefont {Machado}(2007)}]{Machado:2007fr}%
  \BibitemOpen
  \bibfield  {author} {\bibinfo {author} {\bibfnamefont {M.~V.~T.}\
  \bibnamefont {Machado}},\ }\href {\doibase 10.1103/PhysRevD.76.054006}
  {\bibfield  {journal} {\bibinfo  {journal} {Phys. Rev.}\ }\textbf {\bibinfo
  {volume} {D76}},\ \bibinfo {pages} {054006} (\bibinfo {year} {2007})},\
  \Eprint {http://arxiv.org/abs/0705.1005} {arXiv:0705.1005 [hep-ph]}
  \BibitemShut {NoStop}%
\bibitem [{\citenamefont {Khoze}\ \emph {et~al.}(2006)\citenamefont {Khoze},
  \citenamefont {Martin},\ and\ \citenamefont {Ryskin}}]{Khoze:2006uj}%
  \BibitemOpen
  \bibfield  {author} {\bibinfo {author} {\bibfnamefont {V.~A.}\ \bibnamefont
  {Khoze}}, \bibinfo {author} {\bibfnamefont {A.~D.}\ \bibnamefont {Martin}}, \
  and\ \bibinfo {author} {\bibfnamefont {M.~G.}\ \bibnamefont {Ryskin}},\
  }\href {\doibase 10.1088/1126-6708/2006/05/036} {\bibfield  {journal}
  {\bibinfo  {journal} {JHEP}\ }\textbf {\bibinfo {volume} {05}},\ \bibinfo
  {pages} {036} (\bibinfo {year} {2006})},\ \Eprint
  {http://arxiv.org/abs/hep-ph/0602247} {arXiv:hep-ph/0602247} \BibitemShut
  {NoStop}%
\bibitem [{\citenamefont {Albrow}\ \emph
  {et~al.}(2009{\natexlab{a}})\citenamefont {Albrow} \emph
  {et~al.}}]{Albrow:2008pn}%
  \BibitemOpen
  \bibfield  {author} {\bibinfo {author} {\bibfnamefont {M.~G.}\ \bibnamefont
  {Albrow}} \emph {et~al.} (\bibinfo {collaboration} {FP420 R and D}),\ }\href
  {\doibase 10.1088/1748-0221/4/10/T10001} {\bibfield  {journal} {\bibinfo
  {journal} {JINST}\ }\textbf {\bibinfo {volume} {4}},\ \bibinfo {pages}
  {T10001} (\bibinfo {year} {2009}{\natexlab{a}})},\ \Eprint
  {http://arxiv.org/abs/0806.0302} {arXiv:0806.0302 [hep-ex]} \BibitemShut
  {NoStop}%
\bibitem [{\citenamefont {Bonnet}\ \emph {et~al.}(2007)\citenamefont {Bonnet},
  \citenamefont {Pierzchala}, \citenamefont {Piotrzkowski},\ and\ \citenamefont
  {Rodeghiero}}]{Bonnet:2007pw}%
  \BibitemOpen
  \bibfield  {author} {\bibinfo {author} {\bibfnamefont {L.}~\bibnamefont
  {Bonnet}}, \bibinfo {author} {\bibfnamefont {T.}~\bibnamefont {Pierzchala}},
  \bibinfo {author} {\bibfnamefont {K.}~\bibnamefont {Piotrzkowski}}, \ and\
  \bibinfo {author} {\bibfnamefont {P.}~\bibnamefont {Rodeghiero}},\
  }\href@noop {} {\bibfield  {journal} {\bibinfo  {journal} {Acta Phys.
  Polon.}\ }\textbf {\bibinfo {volume} {B38}},\ \bibinfo {pages} {477}
  (\bibinfo {year} {2007})},\ \Eprint {http://arxiv.org/abs/hep-ph/0703320}
  {arXiv:hep-ph/0703320} \BibitemShut {NoStop}%
\bibitem [{\citenamefont {Roland}(2010)}]{Roland:2010ch}%
  \BibitemOpen
  \bibfield  {author} {\bibinfo {author} {\bibfnamefont {B.}~\bibnamefont
  {Roland}},\ }\href@noop {} {\enquote {\bibinfo {title} {{Forward Physics
  Capabilities of CMS with the CASTOR and ZDC detectors}},}\ } (\bibinfo {year}
  {2010}),\ \Eprint {http://arxiv.org/abs/1008.0592} {arXiv:1008.0592
  [physics.ins-det]} \BibitemShut {NoStop}%
\bibitem [{\citenamefont {Albrow}\ \emph
  {et~al.}(2009{\natexlab{b}})\citenamefont {Albrow} \emph
  {et~al.}}]{Albrow:2008az}%
  \BibitemOpen
  \bibfield  {author} {\bibinfo {author} {\bibfnamefont {M.}~\bibnamefont
  {Albrow}} \emph {et~al.} (\bibinfo {collaboration} {USCMS}),\ }\href
  {\doibase 10.1088/1748-0221/4/10/P10001} {\bibfield  {journal} {\bibinfo
  {journal} {JINST}\ }\textbf {\bibinfo {volume} {4}},\ \bibinfo {pages}
  {P10001} (\bibinfo {year} {2009}{\natexlab{b}})},\ \Eprint
  {http://arxiv.org/abs/0811.0120} {arXiv:0811.0120 [hep-ex]} \BibitemShut
  {NoStop}%
\bibitem [{\citenamefont {Lamsa}\ and\ \citenamefont
  {Orava}(2009)}]{Lamsa:2009ej}%
  \BibitemOpen
  \bibfield  {author} {\bibinfo {author} {\bibfnamefont {J.~W.}\ \bibnamefont
  {Lamsa}}\ and\ \bibinfo {author} {\bibfnamefont {R.}~\bibnamefont {Orava}},\
  }\href {\doibase 10.1088/1748-0221/4/11/P11019} {\bibfield  {journal}
  {\bibinfo  {journal} {JINST}\ }\textbf {\bibinfo {volume} {4}},\ \bibinfo
  {pages} {P11019} (\bibinfo {year} {2009})},\ \Eprint
  {http://arxiv.org/abs/0907.3847} {arXiv:0907.3847 [physics.acc-ph]}
  \BibitemShut {NoStop}%
\bibitem [{\citenamefont {Rainwater}\ \emph {et~al.}(2002)\citenamefont
  {Rainwater}, \citenamefont {Spira},\ and\ \citenamefont
  {Zeppenfeld}}]{Rainwater:2002hm}%
  \BibitemOpen
  \bibfield  {author} {\bibinfo {author} {\bibfnamefont {D.~L.}\ \bibnamefont
  {Rainwater}}, \bibinfo {author} {\bibfnamefont {M.}~\bibnamefont {Spira}}, \
  and\ \bibinfo {author} {\bibfnamefont {D.}~\bibnamefont {Zeppenfeld}},\
  }\href@noop {} {\enquote {\bibinfo {title} {{Higgs boson production at hadron
  colliders: Signal and background processes}},}\ } (\bibinfo {year} {2002}),\
  \Eprint {http://arxiv.org/abs/hep-ph/0203187} {arXiv:hep-ph/0203187}
  \BibitemShut {NoStop}%
\bibitem [{\citenamefont {Khoze}\ \emph {et~al.}(2002)\citenamefont {Khoze},
  \citenamefont {Martin},\ and\ \citenamefont {Ryskin}}]{Khoze:2001xm}%
  \BibitemOpen
  \bibfield  {author} {\bibinfo {author} {\bibfnamefont {V.~A.}\ \bibnamefont
  {Khoze}}, \bibinfo {author} {\bibfnamefont {A.~D.}\ \bibnamefont {Martin}}, \
  and\ \bibinfo {author} {\bibfnamefont {M.~G.}\ \bibnamefont {Ryskin}},\
  }\href {\doibase 10.1007/s100520100884} {\bibfield  {journal} {\bibinfo
  {journal} {Eur. Phys. J.}\ }\textbf {\bibinfo {volume} {C23}},\ \bibinfo
  {pages} {311} (\bibinfo {year} {2002})},\ \Eprint
  {http://arxiv.org/abs/hep-ph/0111078} {arXiv:hep-ph/0111078} \BibitemShut
  {NoStop}%
\bibitem [{\citenamefont {d'Enterria}\ and\ \citenamefont
  {Lansberg}(2010)}]{d'Enterria:2009er}%
  \BibitemOpen
  \bibfield  {author} {\bibinfo {author} {\bibfnamefont {D.}~\bibnamefont
  {d'Enterria}}\ and\ \bibinfo {author} {\bibfnamefont {J.-P.}\ \bibnamefont
  {Lansberg}},\ }\href {\doibase 10.1103/PhysRevD.81.014004} {\bibfield
  {journal} {\bibinfo  {journal} {Phys. Rev.}\ }\textbf {\bibinfo {volume}
  {D81}},\ \bibinfo {pages} {014004} (\bibinfo {year} {2010})},\ \Eprint
  {http://arxiv.org/abs/0909.3047} {arXiv:0909.3047 [hep-ph]} \BibitemShut
  {NoStop}%
\bibitem [{\citenamefont {Miller}(2007)}]{Miller:2007pc}%
  \BibitemOpen
  \bibfield  {author} {\bibinfo {author} {\bibfnamefont {J.~S.}\ \bibnamefont
  {Miller}},\ }\href@noop {} {\enquote {\bibinfo {title} {{Electromagnetic
  Higgs production}},}\ } (\bibinfo {year} {2007}),\ \Eprint
  {http://arxiv.org/abs/0704.1985} {arXiv:0704.1985 [hep-ph]} \BibitemShut
  {NoStop}%
\bibitem [{\citenamefont {Khoze}\ \emph {et~al.}(1997)\citenamefont {Khoze},
  \citenamefont {Martin},\ and\ \citenamefont {Ryskin}}]{Khoze:1997dr}%
  \BibitemOpen
  \bibfield  {author} {\bibinfo {author} {\bibfnamefont {V.~A.}\ \bibnamefont
  {Khoze}}, \bibinfo {author} {\bibfnamefont {A.~D.}\ \bibnamefont {Martin}}, \
  and\ \bibinfo {author} {\bibfnamefont {M.~G.}\ \bibnamefont {Ryskin}},\
  }\href {\doibase 10.1016/S0370-2693(97)00426-7} {\bibfield  {journal}
  {\bibinfo  {journal} {Phys. Lett.}\ }\textbf {\bibinfo {volume} {B401}},\
  \bibinfo {pages} {330} (\bibinfo {year} {1997})},\ \Eprint
  {http://arxiv.org/abs/hep-ph/9701419} {arXiv:hep-ph/9701419} \BibitemShut
  {NoStop}%
\bibitem [{\citenamefont {Gay~Ducati}\ and\ \citenamefont
  {Silveira}(2010)}]{GayDucati:2010xi}%
  \BibitemOpen
  \bibfield  {author} {\bibinfo {author} {\bibfnamefont {M.~B.}\ \bibnamefont
  {Gay~Ducati}}\ and\ \bibinfo {author} {\bibfnamefont {G.~G.}\ \bibnamefont
  {Silveira}},\ }\href {\doibase 10.1103/PhysRevD.82.073004} {\bibfield
  {journal} {\bibinfo  {journal} {Phys. Rev.}\ }\textbf {\bibinfo {volume}
  {D82}},\ \bibinfo {pages} {073004} (\bibinfo {year} {2010})},\ \Eprint
  {http://arxiv.org/abs/1007.1182} {arXiv:1007.1182 [hep-ph]} \BibitemShut
  {NoStop}%
\end{thebibliography}
\end{document}